\documentclass[fleqn,usenatbib]{mnras}

\usepackage{newtxtext,newtxmath}

\usepackage[T1]{fontenc}

\DeclareRobustCommand{\VAN}[3]{#2}
\let\VANthebibliography\thebibliography
\def\thebibliography{\DeclareRobustCommand{\VAN}[3]{##3}\VANthebibliography}

\usepackage{graphicx}	
\usepackage{amsmath}	

\newcommand{\keckgdc}{K09}
\newcommand{\keckjwa}{K12}
\newcommand{\keckjwb}{K14}
\newcommand{\keckjwc}{K15}
\newcommand{\almagdc}{A02}
\newcommand{\almajwext}{A03}
\newcommand{\almajwcomp}{A04}
\newcommand{\keckbs}{K13}

\title[ALMA and Keck analysis of Fomalhaut field sources]{ALMA and Keck analysis of Fomalhaut field sources: JWST's Great Dust Cloud is a background object}
\author[G. M. Kennedy, J. B. Lovell, P. Kalas \& M. P. Fitzgerald]{
Grant M. Kennedy$^{1}$\thanks{E-mail: g.kennedy@warwick.ac.uk}, Joshua B. Lovell$^{2}$, Paul Kalas$^{3,4,5}$ \& Michael P. Fitzgerald$^{6}$
\\
$^{1}$Department of Physics and Centre for Exoplanets and Habitability, University of Warwick, Gibbet Hill Road, Coventry CV4 7AL, UK \\
$^{2}$Center for Astrophysics, Harvard \& Smithsonian, 60 Garden Street, Cambridge, MA 02138-1516, USA \\
$^{3}$ University of California, Berkeley, 501 Campbell Hall, 3411 Berkeley, CA 94720-3411, USA \\
$^4$Institute of Astrophysics, FORTH, GR-71110 Heraklion, Greece\\
$^{5}$SETI Institute, Carl Sagan Center, 189 Bernardo Ave.,  Mountain View, CA 94043, USA\\
$^6$Department of Physics and Astronomy, University of California, Los Angeles, 430 Portola Plaza, Box 951547, Los Angeles, CA 90095-1547, USA
}
\date{Accepted 2023 July 4. Received 2023 July 4; in original form 2023 May 17}

\pubyear{2023}

\begin{document}
\label{firstpage}
\pagerange{\pageref{firstpage}--\pageref{lastpage}}
\maketitle

\begin{abstract}
At 7.7\,pc, the A-type star Fomalhaut hosts a bright debris disk with multiple radial components. The disk is eccentric and misaligned, strongly suggesting that it is sculpted by interaction with one or more planets. Compact sources are now being detected with JWST, suggesting that new planet detections may be imminent. However, to confirm such sources as companions, common proper motion with the star must be established, as with unprecedented sensitivity comes a high probability that planet candidates are actually background objects. Here, ALMA and Keck observations of Fomalhaut are found to show significant emission at the same sky location as multiple compact sources in JWST MIRI coronagraphic observations, one of which has been dubbed the "Great Dust Cloud" because it lies within the outer belt. Since the ground-based data were obtained between 6 to 18 years prior to the JWST observations, these compact sources are unlikely to be common proper motion companions to Fomalhaut. More generally, this work illustrates that images collected at a range of wavelengths can be valuable for rejecting planet candidates uncovered via direct imaging with JWST.
\end{abstract}




\section{Introduction}

Fomalhaut is a bright ($m_V$=1.16 mag) nearby (7.7\,pc) A-type star known to host a bright circumstellar dust disk \citep{1986ASSL..124...61G}. The system has been studied in detail for decades, yielding optical images of dust-scattered light and a compact dusty source \citep{2005Natur.435.1067K,2008Sci...322.1345K}, and far-infrared and mm-wave images of dust thermal emission \citep[e.g.][]{2003ApJ...582.1141H,2004ApJS..154..458S,2017ApJ...842....8M} and gas \citep{2017ApJ...842....9M}.

A key question for Fomalhaut, as with all stellar systems, is ``where are the planets?'' While in the past such a question might have seemed to be optimistic, the \emph{Kepler} mission \citep{2003SPIE.4854..129B} has proven beyond doubt that planets are common. In addition, Fomalhaut's bright debris ring is eccentric \citep{2005Natur.435.1067K}, for which the simplest explanation is that the ring material is perturbed by an eccentric planet \citep{1999ApJ...527..918W}. Thus, this system has been, and will continue to be, a prime target in the ongoing search for planets that can be directly imaged.

The possibility of planets has also been highlighted by new images presented by \citet{2023NatAs.tmp...93G} from the Mid-Infrared Instrument \citep[MIRI;][]{2008SPIE.7010E..0TW} on the James Webb Space Telescope \citep[JWST;][]{2006SSRv..123..485G}. These data show that the Fomalhaut disk has additional structure that has not been previously resolved. While an asteroid belt analog (ABA) at $\sim$11\,au has long been known to exist \citep[e.g.][]{2004ApJS..154..458S,2012A&A...540A.125A,2013ApJ...763..118S} and is confirmed with JWST, the images reveal a previously unseen intermediate belt at $\sim$90\,au that may be misaligned with the outer belt. The outer belt at $\sim$140\,au is dubbed the Kuiper Belt Analog (KBA) by \citet{2023NatAs.tmp...93G}. With a series of radially nested belts that are separated by gaps, Fomalhaut somewhat resembles an eccentric and misaligned version of the ringed protoplanetary disks \citep[e.g.][]{2015ApJ...808L...3A,2018ApJ...869L..41A}. These continuum radial structures in protoplanetary disks are thought to be caused by a combination of planetary perturbation and dust drift \citep[e.g.][]{2006MNRAS.373.1619R,2012A&A...538A.114P}, so it seems reasonable that systems such as Fomalhaut inherit a similar structure of planets and dusty rings.

Another JWST discovery at 23.0 and 25.5\,$\mu$m is a compact structure that lies west of the star within the KBA, dubbed the ``Great Dust Cloud'' (GDC). This source is 89\,$\mu$Jy at 25.5\,$\mu$m, compared to the star's 2.5\,Jy, so extremely faint. Because it appears in both the 23.0 and 25.5\,$\mu$m data, it is confirmed as a real astrophysical source as opposed to an instrumental or data reduction artifact. The fact that the GDC lies almost perfectly within Fomalhaut's narrow KBA (i.e. a special location in the image) suggests that it is less likely to be a background source, though \citeauthor{2023NatAs.tmp...93G} make clear that with only a single epoch of JWST observation the possibility that the GDC is a background source cannot be ruled out. The idea that dust clumps could be created and reside within a debris disk is not new; \citet{2002MNRAS.334..589W} developed a theory for collisional clumps, and in the case of the Fomalhaut disk, suggested that these may be detectable with ALMA and/or a future large-aperture mid-infrared space telescope (i.e. JWST). 

In this work we aim to test whether the putative dust cloud is indeed associated with Fomalhaut; the system's proper motion of $\mu_\alpha=329.0$\,mas\,yr$^{-1}$ and $\mu_\delta=-164.7$\,mas\,yr$^{-1}$  \citep{2007A&A...474..653V} means that previously detected background sources may now lie behind the KBA, masquerading as dust clouds. We revisit archival ALMA data for this system from 2015--2016, which show a source that lies at the sky location of the GDC. We validate our finding that the GDC is a background object with previously unpublished Keck adaptive optics (AO) data of Fomalhaut obtained at multiple epochs obtained 11--18 years before the JWST observations were made.

\section{Data}

The ALMA 1.3\,mm continuum data have appeared in two publications \citep{2017ApJ...842....8M,2020RSOS....700063K}, and the processed JWST data (GTO-1193; PI Beichman) have been made available in FITS format by \citet{2023NatAs.tmp...93G}.\footnote{\href{https://github.com/merope82/Fomalhaut}{https://github.com/merope82/Fomalhaut}} Readers can refer to those publications for details on their reduction and calibration procedures, but we will review the most relevant points here. A portion of the Keck data were used in \citet{2008Sci...322.1345K} where the objective was to detect Fomalhaut b rather than make a map of background sources. In the present analysis we include subsequent epochs of Keck $H$-band observations made through 2011 (Table~\ref{sec:keckobs}).

\subsection{ALMA}

Atacama Large Millimetre/Submillimetre Array (ALMA) observations of the entire Fomalhaut ring were obtained in late 2015 December and early 2016 January in Band\,6 (1.3\,mm) \citep{2017ApJ...842....8M}. They are a mosaic of seven pointings that can be combined to make a single large image. Being an interferometer, ALMA's sky sensitivity is dictated by its primary beam, with a full-width at half-maximum of approximately 22\,\arcsec; while the sensitivity in uncorrected images appears uniform, the signal is attenuated by the primary beam and is in fact significantly noisier farther from the image center. Thus, to be detected, sources farther from the image center must be brighter. The spatial resolution of ALMA is set by relative antennae positioning and spacing at the time of observation, which for this data provide a beam with 1.6${\times}$1.2\,\arcsec~resolution. The clean images show that the star is strongly detected and lies within a few tens of mas of the expected coordinates at the observation epoch.

These data were also presented in \citet{2020RSOS....700063K} who analyzed the KBA width by creating a model fit to the observations (the ABA and intermediate belt were not detected). To demonstrate the fidelity of the model, their Fig. 6 shows the difference image after Fomalhaut's dust belt is subtracted by the model. Black contours identified residual emission in the entire field greater than 3$\sigma$ above the noise, where $\sigma=13$\,$\mu$Jy. In the present analysis we will discuss residual emission sources 4$\sigma$ above the noise.

\subsection{Keck}

\begin{table}
\centering
\caption{Keck $H$-band observing log.}\label{tab:keckobs}
\begin{tabular}{ccccc}
UT & MJD & N$_{\rm exp}$ & t$_{\rm int}$ & $\Delta$P.A. \\
 &  (midpoint) &  & (s) & (deg) \\
\hline
2005-07-17 & 53568.59 & 117 & \phantom{0}3790.0 & 41.1\\
2005-07-27 & 53578.54 & 144 & \phantom{0}4320.0 & 38.4\\
2005-07-28 & 53579.55 & 163 & \phantom{0}4890.0 & 44.8\\
2005-10-21 & 53664.31 & 174 & \phantom{0}5197.3 & 44.8\\
2010-07-02 & 55379.55 & 359 & 10770.0 & 69.2\\
2010-07-03 & 55380.55 & 404 & 12105.0 & 71.8\\
2010-11-11 & 55511.26 & 217 & \phantom{0}5610.0 & 41.1\\
2011-07-12 & 55754.57 & 138 & \phantom{0}2028.6 & 27.5\\
2011-07-14 & 55756.55 & 215 & \phantom{0}9178.0 & 66.2\\
\label{sec:keckobs}
\end{tabular}
\end{table}

\begin{table}
\caption{IDs and positions of sources (J2000).}\label{tab:sources}
\begin{tabular}{lccccc}
ID & RA & Dec & JWST & Keck & ALMA \\
\hline
K01 & 22:57:37.44 & -29:37:13.86 &  & y &  \\
K02 & 22:57:37.57 & -29:37:22.51 &  & y &  \\
K03 & 22:57:37.61 & -29:37:31.21 &  & y &  \\
K04 & 22:57:38.08 & -29:37:32.25 & y & y &  \\
K05 & 22:57:38.16 & -29:37:17.86 &  & y &  \\
K06 & 22:57:38.24 & -29:37:12.97 &  & y &  \\
A01 & 22:57:38.31 & -29:37:05.52 &  &  & y \\
K07 & 22:57:38.79 & -29:37:34.25 &  & y &  \\
D01* & 22:57:38.88 & -29:37:04.84 &  &  & y \\
K08 & 22:57:39.09 & -29:37:44.15 & y & y &  \\
{\bf A02*} & 22:57:39.15 & -29:37:25.59 & y & y & y \\
{\bf K09*} & 22:57:39.15 & -29:37:26.18 & y & y & y \\
{\bf GDC-25*} & 22:57:39.16 & -29:37:26.22 & y & y & y \\
{\bf GDC-23*} & 22:57:39.18 & -29:37:26.18 & y & y & y \\
K10 & 22:57:39.34 & -29:37:04.03 &  & y &  \\
K11* & 22:57:39.34 & -29:37:05.20 &  & y &  \\
K12* & 22:57:39.80 & -29:37:28.79 & y & y &  \\
K13* & 22:57:40.07 & -29:37:36.81 &  & y &  \\
D02* & 22:57:40.15 & -29:37:35.12 &  &  & y \\
K14* & 22:57:40.17 & -29:37:29.93 & y & y &  \\
K15* & 22:57:40.41 & -29:37:34.53 & y & y &  \\
K16 & 22:57:40.62 & -29:37:34.19 &  & y &  \\
A03, K17 & 22:57:40.74 & -29:37:32.43 & y & y & y \\
K18 & 22:57:41.02 & -29:37:08.22 &  & y &  \\
K19 & 22:57:41.04 & -29:37:25.58 & y & y &  \\
A04 & 22:57:41.45 & -29:37:21.54 & y &  & y \\
\hline
\end{tabular}
Notes: The last three columns indicate which instrument sources were detected by (JWST refers to either 23 or 25.5\,$\mu$m). Starred sources lie on or interior to the KBA. The IDs \almagdc, \keckgdc, GDC-23 (JWST 23.0\,$\mu$m), and GDC-25 (JWST 25.5\,$\mu$m) are all the same source. Likewise A03 and K17 are the same source.
\end{table}

At Keck II, Fomalhaut was imaged with AO and NIRC2 \citep{2000PASP..112..315W} using its widest field of 40$\times$40\,\arcsec~(40\,mas\,pix$^{-1}$) at 1.6\,$\mu$m. The star is sufficiently bright to serve as its own wavefront reference with a 1\,kHz readout rate for the wavefront sensor, along with a neutral density filter in the wavefront sensing optical path. As a southern target, we generally attempted to observe it from the time it reached a reasonable elevation angle until the telescope could no longer track it due to the Nasmyth deck pointing limit. The image rotator located at the Nasmyth focus was placed in vertical angle mode, which stabilizes the orientation of the telescope pupil relative to the detector, and enables angular differential imaging (ADI). Due to the brightness of Fomalhaut ($m_R$=1.11\,mag, $m_H$=1.05\,mag), we used a large (2.0\,\arcsec~diameter) focal plane mask in order to prevent electronic artifacts in detector readout. We chose exposure times to allow for deep integration, resulting in saturation and nonlinearity in a region extending radially beyond the edge of the focal plane mask. The size of this region varied with conditions and the degree of AO correction. With good conditions and AO performance, the PSF full-width at half-maximum is $\sim$60\,mas (undersampled relative to the Nyquist limit).

Observations were classically scheduled and weather conditions varied. Cloud cover and the degree of atmospheric turbulence had significant impacts on the sensitivity achieved at any given epoch. Individual exposures were reduced using standard techniques to correct for electronic biases, flat field (with the focal plane mask in place), detector nonlinearity, and field distortion.

To subtract Fomalhaut's point spread function (PSF) we used the LOCI technique on these ADI data \citep{2007ApJ...660..770L}. For a given exposure, this method combines other images in the sequence to construct a reference PSF. The data for each night were processed separately.

The data obtained throughout each night and from night-to-night are subject to variable extinction. The PSF of all sources will also change over time with variations in the on-axis AO correction that is sensitive to changes in the vertical profiles of atmospheric turbulence strengths and wind velocities, quasi-static aberrations, etc. In addition, the time variability of high-altitude atmospheric turbulence introduces fluctuations in the structure of the off-axis AO PSF. 

\begin{figure*}
    \hspace{-0.5cm}
    \includegraphics[height=0.5\textwidth]{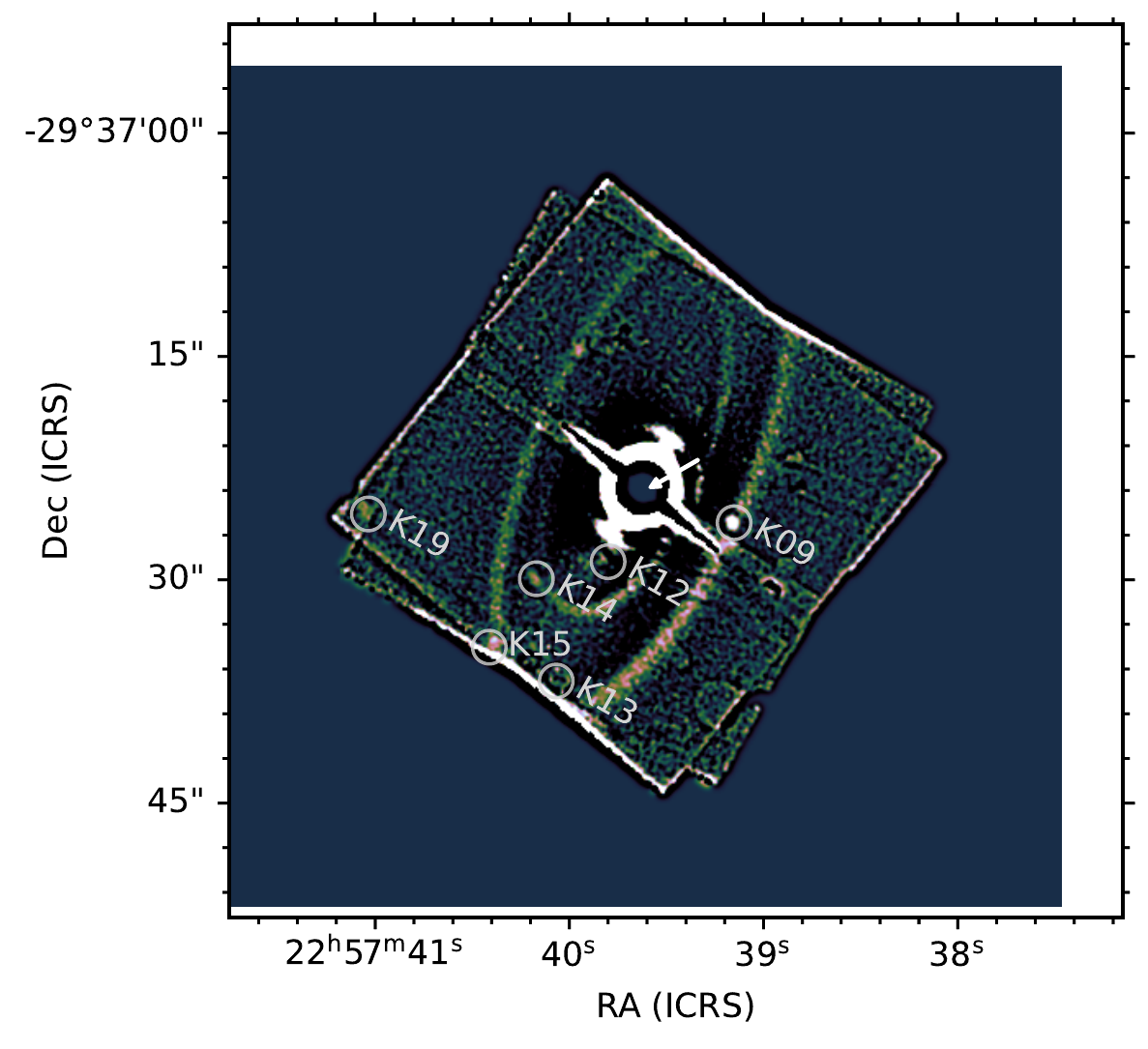}
    \includegraphics[height=0.5\textwidth]{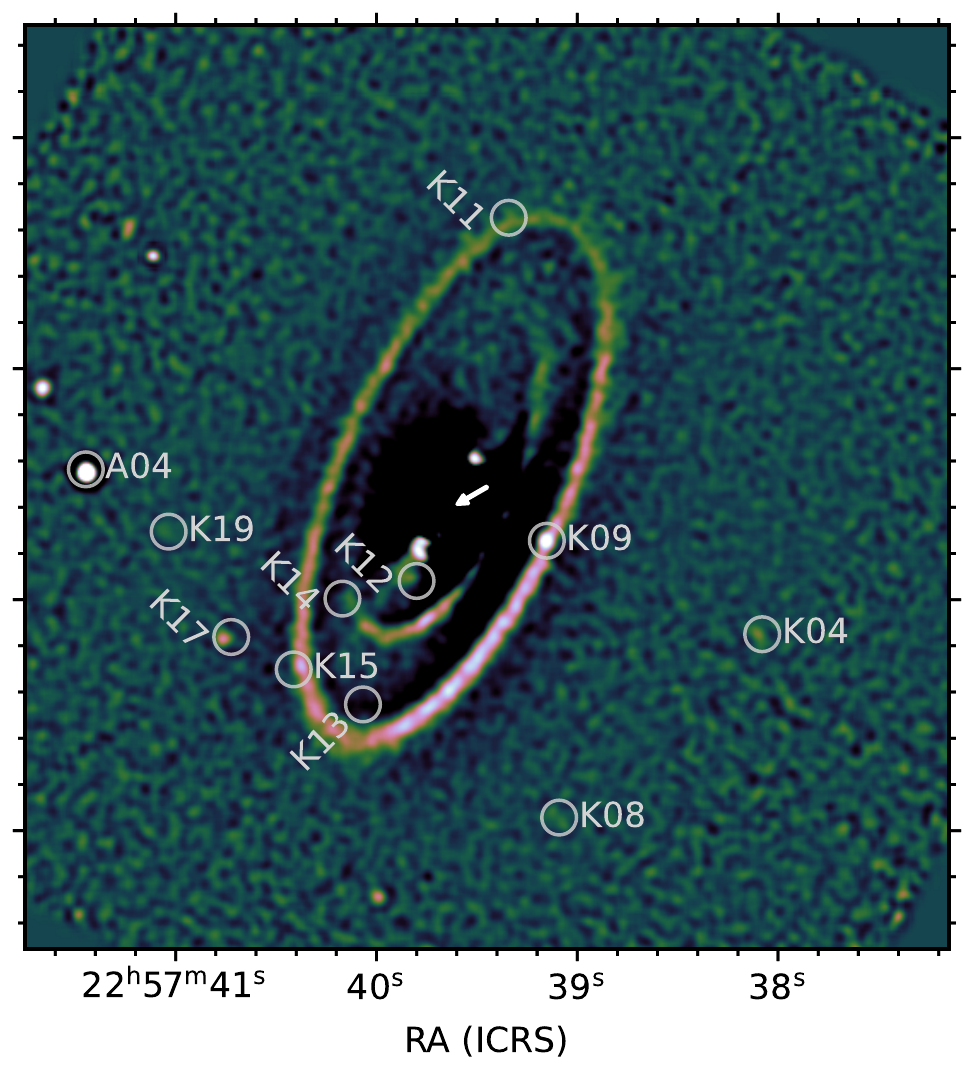}
    \caption{Fomalhaut's belt features after high-pass filtering the 23.0 (left panel) and 25.5\,$\mu$m data (right panel). Both images have the same size and center. Arrows show the stellar proper motion since 2005 July (left panel) and 2015 December (right panel). Keck sources and \almajwcomp~are circled and labelled. The circles are approximately 200\,mas in diameter. The 23.0\,$\mu$m image in the left panel has been moved 0.3\,\arcsec~W to account for a difference in the apparent GDC location between 25.5 and 23.0\,$\mu$m.}
    \label{fig:jwst2325}
\end{figure*}

Since it is detectable in individual exposures, we use the time variability of the bright star southeast of Fomalhaut (\keckbs, see Table~\ref{tab:sources}) to track changes in the AO performance and extinction. This allows us to combine images during each night and across nights by assigning a weight factor that incorporates the signal recovered from the background star. We also incorporate a measure of the PSF-subtraction fidelity, by using an estimate of the subtraction uncertainty derived from the $\chi^2$ measure of the LOCI optimization. We use a robust combination method that incorporates the above weights. All images across all nights of observations in a given band are combined after registration to the celestial reference frame, thus co-adding the signal from background sources.  

Any sources that have a common proper motion with Fomalhaut would appear as triplet images in the combined frame corresponding to the 2005, 2010, and 2011 epochs of observation (there are no such sources detected). All other sources are background objects which we can further verify by blinking the data from the three epochs and using additional versions of the images that are binned and smoothed. When these data from different epochs are registered to the position of Fomalhaut, blinking shows the background sources moving along an axis oriented at PA = 116.59$\degr$ (Fomalhaut's proper motion vector) relative to any local noise. Thus, even though source \keckgdc~lies close to the central speckle halo, we confirmed it as a real astrophysical object using this method eight years before JWST was launched. Therefore \keckgdc~was identified independently from the JWST results, so as with the ALMA sources there is no confirmation bias. 

In Table~\ref{tab:sources} we provide the centroid positions of background sources detected and validated by our Keck program. We adopt a conservative position uncertainty estimate of 80\,mas to account for the systematic and random errors involved in determining the position of Fomalhaut, the true north orientation, uncorrected geometric distortions, and the centroiding measurement itself that depends on the signal-to-noise ratio (SNR) and morphology of each source. Several of our sources have been previously noted as background objects by \citet{2013ApJ...777L...6C}; their source 1 is K13, 2=K04, 3=K07, 4=K10, 5=K11, 6=K14, and 7=K12. With the exception of K09 below, we have not presented magnitudes as many sources are resolved with a size of a few tens of mas and hence may be affected by the self-subtraction that is inherent in LOCI images. As an approximate reference, K13 is Gaia~DR3~6606098685561921280 with a $G$ magnitude of $18.22 \pm 0.02$ and a $G_{BP}-G_{RP}$ colour consistent with zero, suggesting that the $H$-band magnitude is similar.

\subsection{JWST}

JWST's MIRI field of view (FOV) at 25.5\,$\mu$m is large: 56$\times$56\,\arcsec~in these Fomalhaut observations using the ``BRIGHTSKY'' subarray. MIRI can capture the entire Fomalhaut disk in a single FOV, retaining high sensitivity to the dusty outer regions and to compact background sources that may be significantly separated from Fomalhaut's disk, as well as any that could be coincident with it. The 23.0\,$\mu$m data have a smaller FOV (30$\times$30\,\arcsec) that do not capture the southeast and northwest portions of Fomalhaut's KBA. The angular resolution of MIRI is higher than the ALMA observations, approximately 0.5\,\arcsec. We do not carry out a complete search for point sources in the MIRI fields, but identify those that are also detected with ALMA and/or Keck. The GDC is not visible at 15.5\,$\mu$m, so we do not consider those data here.

To assist in measuring the positions of features in the KBA and intermediate belt, we high-pass filtered the MIRI data.  We smoothed each image with a Gaussian ($\sigma$=2\,pixels) and subtracted the smoothed images from the originals. The difference images are smoothed by a Gaussian ($\sigma$=1\,pixel at 23.0\,$\mu$m, $\sigma$=3\,pixels at 25.5\,$\mu$m) and shown in Figure \ref{fig:jwst2325}, with our measurements of GDC positions listed as GDC-23 and GDC-25 in Table \ref{tab:sources}.

Our centroid measurements for GDC reveal that the 23.0\,$\mu$m position lies 0.3\,\arcsec~east of its 25.5\,$\mu$m position. Blinking the two JWST images when registered to the world coordinate system shows that the intermediate belt and KBA also have the same shift between images. This shift is evidence for a systematic error in determining the location of Fomalhaut on the detector at each wavelength or some other factor in the data processing that establishes the celestial reference frame. Our analysis requires a comparison of source positions in the time domain at different wavelengths. However, since this is a relatively sparse field, source confusion is not a significant problem and the 0.3\,\arcsec~position uncertainty for JWST astrometry is unlikely to change our conclusions.
The GDC position could serve as a guide since Table \ref{tab:sources} shows that the 25.5\,$\mu$m position instead of the 23.0\,$\mu$m position agrees more closely with
both the ALMA and Keck data.

The JWST observations were carried out on 2022 October 22, nearly seven years after the ALMA observations and more than 11 years after the last Keck observation. Given its southeastward proper motion of 367.9\,mas/yr, Fomalhaut at the JWST epoch has moved 2.6\,\arcsec~away from the ALMA epoch and 4.1\,\arcsec~away from the last Keck epoch.
In contrast, background sources typically have negligible proper motions, and so would remain static on the sky plane relative to Fomalhaut. Given these high angular resolution observations with Keck, ALMA and JWST, the background sources are easily distinguishable between observations.

\section{Results}

We now compare the ALMA and Keck data to JWST, focusing on the locations of compact sources.

\subsection{Keck}

\begin{figure}
    \hspace{-0.6cm}
    \includegraphics[height=0.48\textwidth]{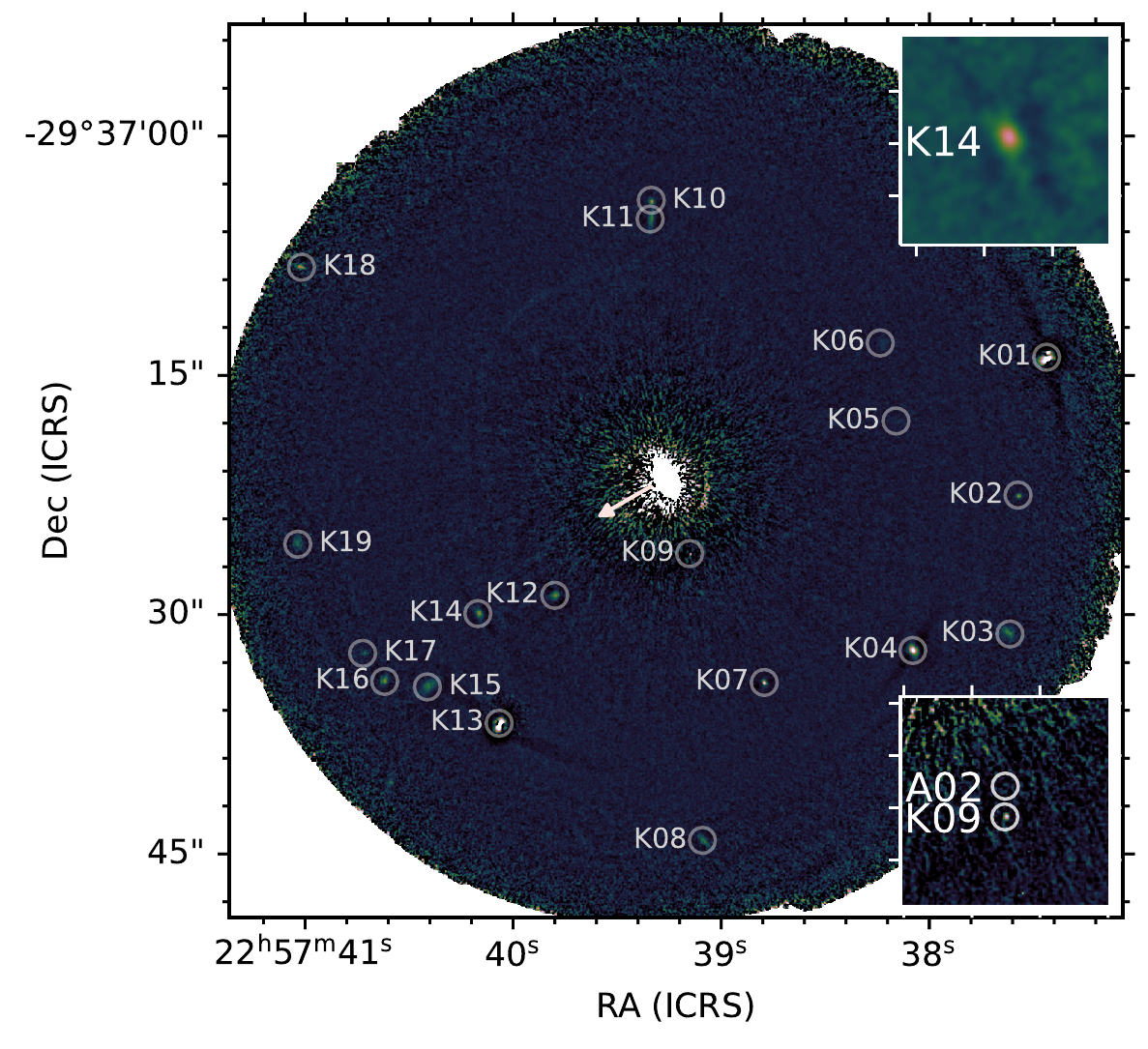}
    \caption{2011 Keck $H$-band image with a log stretch and background sources circled and labelled. The arrow shows the proper motion from 2011 July to 2022 October. The circles in the main panel are approximately 150\,mas in diameter. The insets show 4$\times$4\,\arcsec~zooms centered on the GDC and \keckjwb~which is extended along PA$\sim$30$\degr$.}\label{fig:keck}
\end{figure}

The background sources discovered in the multi-epoch Keck data are labeled in Figure \ref{fig:keck} and overlayed on the JWST 23.0 and 25.5\,$\mu$m images in Figure \ref{fig:jwst2325}. Source K09 is coincident with the GDC and detected at SNR=5. At our epochs of observation it appears as a point source at $\sim$5\,\arcsec~projected radius from Fomalhaut. It is possible that K09 is extended but the residual noise at that radius after subtracting Fomalhaut's bright PSF and the effect of self-subtraction from the LOCI algorithm interferes with the detection of diffuse extended structure. Using a Keck/NIRC2 $H$-band zeropoint of 24.2\,mag, K09 is 20.9$\pm$0.3\,mag. However, this estimate is probably fainter than the true magnitude due to self-subtraction from ADI/LOCI processing. K09 is offset 0.59\,\arcsec~S from the ALMA source \almagdc~discussed below but aligned in the right ascension direction. K09 is 0.12\,\arcsec~W and 0.05\,\arcsec~N of the 25.5\,$\mu$m position (GDC-25, Table \ref{tab:sources}). The discrepancy in right ascension is larger between K09 and GDC-23 (see discussion in Section 2.3) but negligible in declination. 

Table~\ref{tab:sources} lists seven more background sources in the Keck data that appear to have counterparts in the JWST data. K04 is a background galaxy because it appears extended in both the Keck data analyzed here as well as the HST/STIS optical data published in \citet{2013ApJ...775...56K}. K08 has a counterpart in the JWST data that is more easily distinguished by binning and smoothing the 25.5 $\mu$m data instead of high-pass filtering.  It has an extended morphology consistent with a background galaxy. The same is true for K17, which is located outside the Keck FOV in several nights of data, but within the FOV in other observations (2010-07-02, 2010-07-03, and 2011-07-14). K17 is co-incident with A03 which was also discussed in \citet{2017ApJ...842....8M} who show an HST/STIS image of the source.
K19 corresponds to emission on the eastern edge of the 23.0$\mu$m frame (Figure \ref{fig:jwst2325}) that is also apparent in the 25.5\,$\mu$m image using binning and smoothing instead of high-pass filtering. It also has an elliptical morphology in the Keck data consistent with a galaxy.

\begin{figure*}
    \hspace{-0.5cm}
    \includegraphics[height=0.49\textwidth]{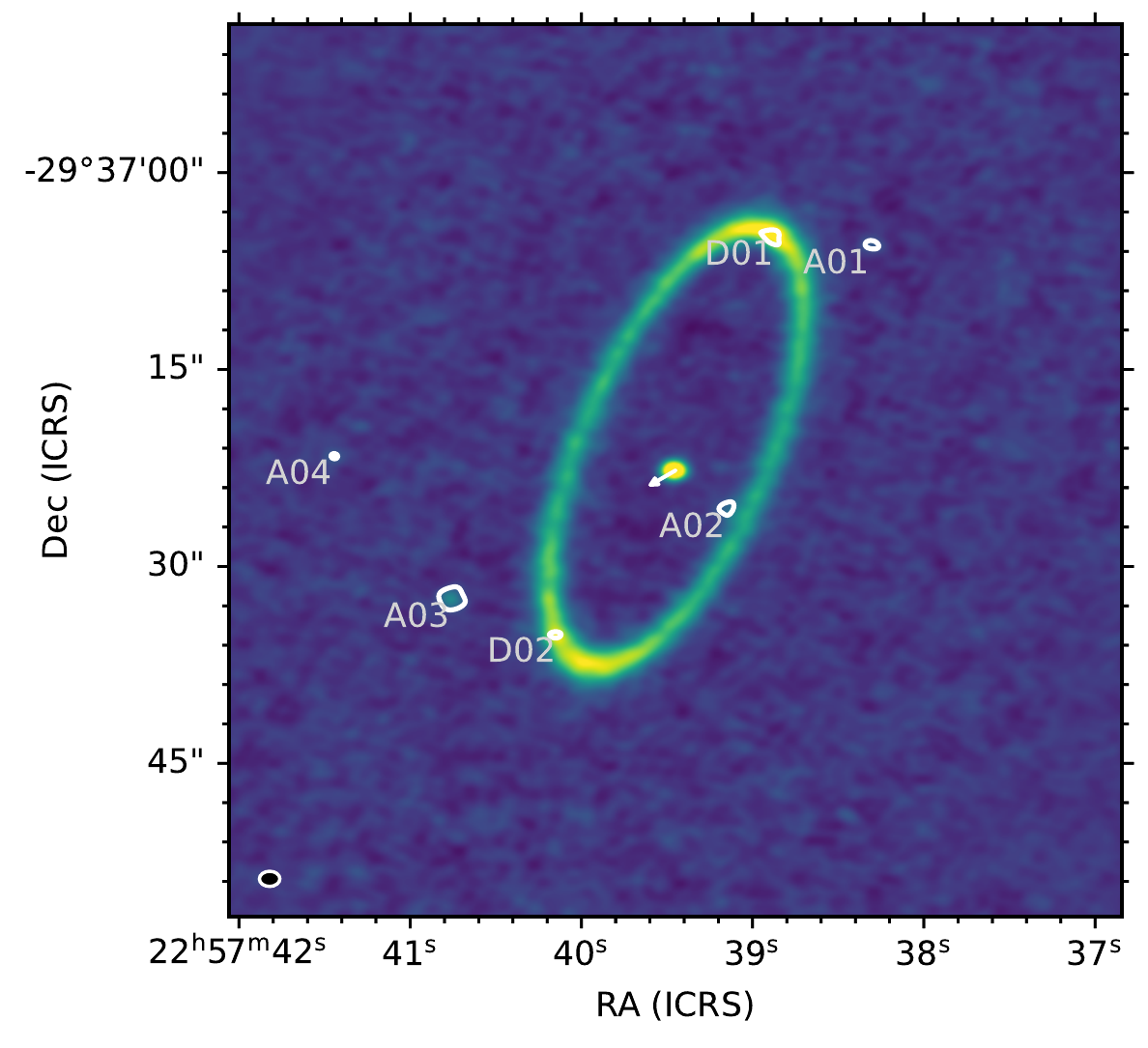}
    \hspace{-.25cm}
    \includegraphics[height=0.49\textwidth]{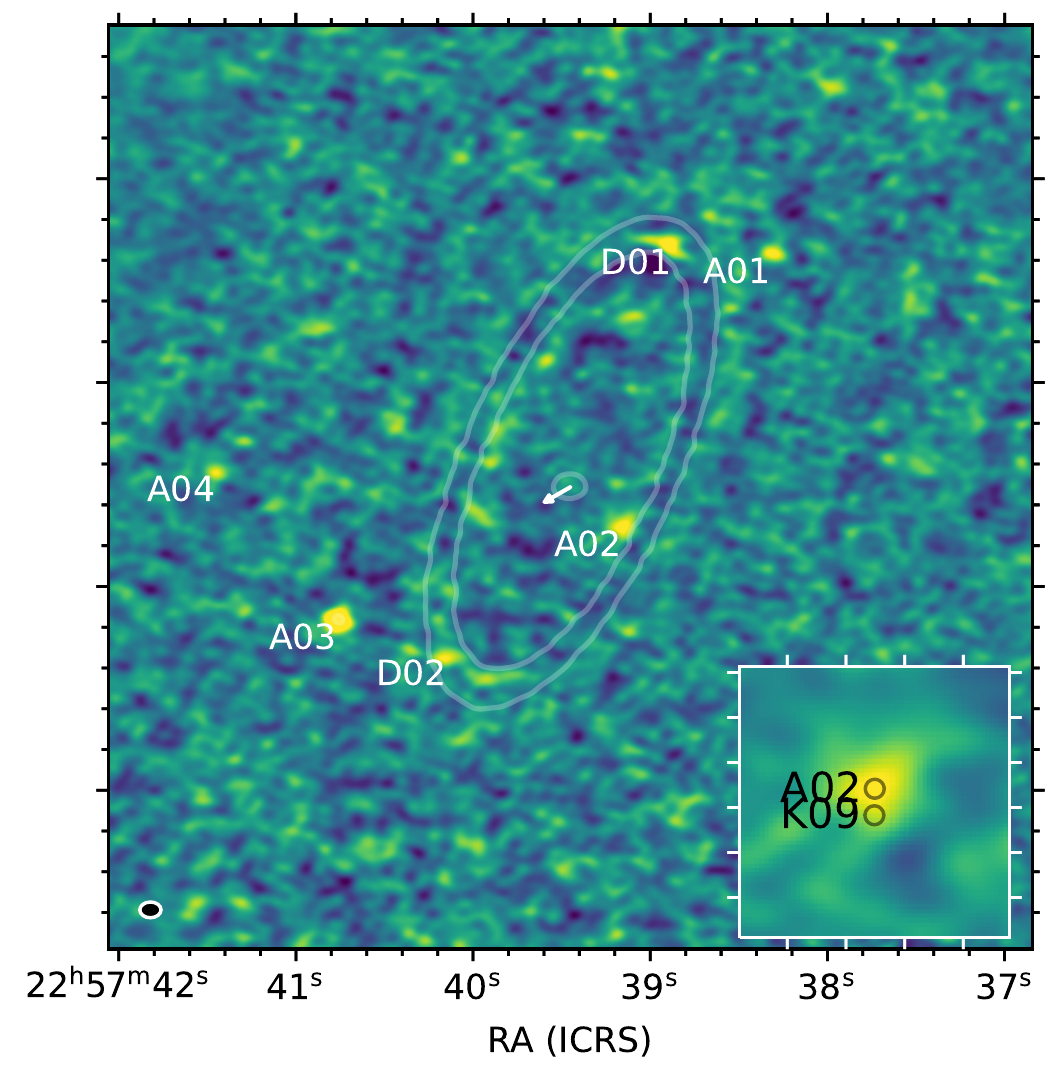}
    \caption{ALMA images. The left panel shows the continuum image with  4\,$\sigma$ residuals as contours, and the right panel shows the residuals with a 10\,$\sigma$ contour from the continuum map. The arrows show the proper motion from 2015 December to 2022 October. The inset shows a 6$\times$6\,\arcsec~zoom of the residual image, showing the sources \almagdc~and \keckgdc~(noting that GDC is at almost exactly the same location as \keckgdc). The beam is shown in the lower left corner of each image. The image is not primary beam corrected, so the flux scale is not uniform across the image. Figures and inset have the same sizes and centers as Fig.~\ref{fig:jwst_alma}.}\label{fig:alma}
\end{figure*}

Three more Keck sources are superimposed on or inside of Fomalhaut's KBA and may influence the interpretations of belt morphologies from JWST data. \keckjwa~may make the ABA appear brighter and more extended to the southeast of the star. \keckjwb~may be misinterpreted as a clump or spiral feature along the intermediate belt, possibly influencing measurements of the belt edges or brightest points needed to derive its position angle and inclination (see also the inset for Fig. \ref{fig:keck} that shows this source is resolved with an elliptical morphology). \keckjwc~is coincident with the KBA and may lead to the same misinterpretation as the GDC. Finally, we note that even though K11 overlaps with the northern part of the belt, it does not appear to have a counterpart in 25.5\,$\mu$m emission.

\subsection{ALMA}

\begin{figure}
    \hspace{-0.5cm}
    \includegraphics[width=0.5\textwidth]{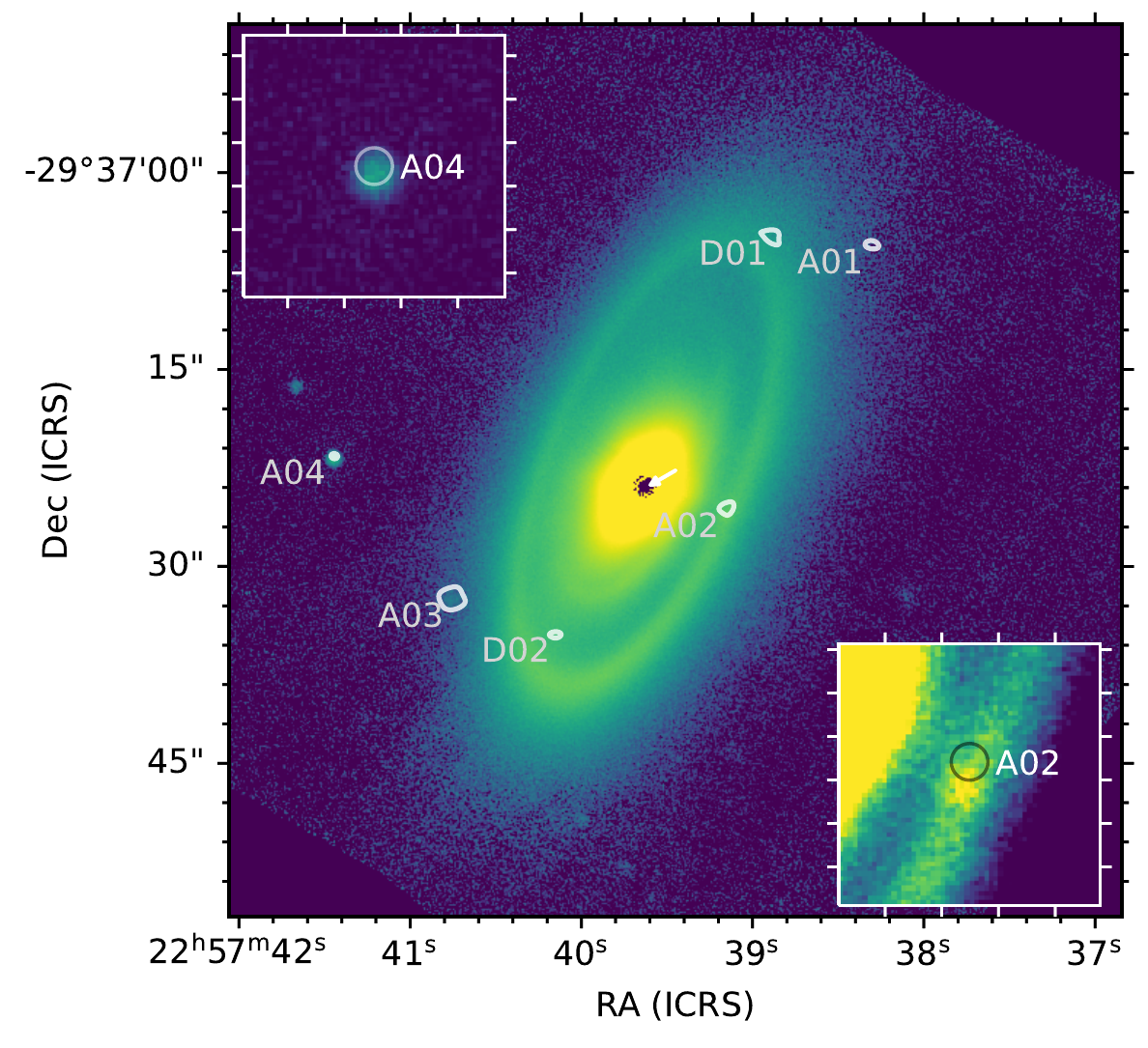}
    \caption{ALMA - JWST comparison. The colour map is the JWST image, the contours are the ALMA residuals, and the labelled sources are significant ALMA residuals. The arrow shows the proper motion from 2015 December to 2022 October. The insets show 6$\times$6\,\,\arcsec~zooms near \almajwcomp~and the GDC. The figure and insets have the same sizes and centers as Fig.~\ref{fig:alma}.}
    \label{fig:jwst_alma}
\end{figure}

The ALMA data, both the image and residuals after model subtraction, are shown in Figure \ref{fig:alma} \citep[reproduced from][]{2020RSOS....700063K}. The sources identified with emission above 4$\sigma$ (where $\sigma=13$\,$\mu$Jy) are given in Table \ref{tab:sources}, none of which are well resolved. The residual labelled \almagdc~only becomes clearly visible once a model is subtracted. For clarity the primary beam correction has not been applied, so the flux scale is not constant across the image. A minor difference relative to the images shown in \citet{2020RSOS....700063K} is that the field of view has been increased to incorporate the source E of the star labelled \almajwcomp. With a beam slightly larger than 1\,\arcsec, and low SNR detections, we conservatively estimate the positional uncertainties to be approximtely 1\,\arcsec.

A number of other significant contours are present in the residuals. Two of these lie near the disk ansae (D01, D02), and are thought to be associated with imperfect modelling of the disk surface brightness, though this conclusion is more secure for D01 as D02 is not exactly at an ansa. As shown by the right panel of Figure \ref{fig:alma}, the model is otherwise a very good representation of the data, so there is no reason to believe that the residual near the GDC is an artefact of the modelling. Aside from these residuals the other three significant residuals lie well away from the disk. While \almajwcomp~appears faint, at this location the primary beam power is about 20\% relative to the image center, so this source is actually about a factor of two brighter than \almajwext. By looking at individual scans we see no evidence that any of the ALMA sources are variable.

As with Keck, there is a clear correspondence between some of the ALMA and JWST sources. Figure \ref{fig:jwst_alma} combines the ALMA contours and JWST 25.5\,$\mu$m image in a single figure. It is clear that \almagdc, \almajwext, and \almajwcomp~correspond to JWST sources, and that these sources are therefore almost certainly background objects that are detected at both wavelengths (with \almagdc~and \almajwcomp~also detected by Keck). In terms of fluxes this conclusion also seems reasonable; these sources have fluxes that are similar within a factor of a few for both ALMA and JWST, so if they are similar types of galaxies, then their detections in both datasets is reasonably expected. The source A01 is not obviously detected with JWST, but may be detected with deeper imaging, or perhaps at other JWST wavelengths (i.e. with NIRCAM).

As shown by the insets in Figure \ref{fig:jwst_alma}, for \almagdc~and \almajwcomp~there are small ($<$0.6\,\arcsec) offsets to the S of the JWST sources relative to the ALMA contours, suggesting that there is a slight offset in coordinate systems. This offset cannot be discerned for \almajwext, but this source is probably marginally resolved by ALMA and detected at low SNR with JWST, so any offset is harder to distinguish. The offset is slightly greater for \almagdc, but the residuals in Figure \ref{fig:alma} show that the location of this source may be biased north by a negative residual to the south.

In summary, as with Keck, the main conclusion from comparing the ALMA and JWST data is that the GDC is a background object.

\subsection{The Great Dust Cloud: star or galaxy?}

In terms of the Fomalhaut system, whether the GDC is a star or galaxy in the background is not important. But JWST will no doubt face similar issues in the future, so we briefly consider the possibilities.

First considering the nature of the GDC, a stellar source is unlikely because it would require a bright infrared excess component; the Keck flux is about 4\,$\mu$Jy and therefore roughly 10--20 times fainter than in the mid-IR and millimeter fluxes, which implies $L_{\rm excess}/L_\star~>~10$\%, a level seen for bright protoplanetary disks. The spectrum is rising from 23 to 25.5\,$\mu$m \citep{2023NatAs.tmp...93G}, which implies the peak is somewhere in the far-IR at a level higher than seen with JWST, which could easily imply $L_{\rm excess}/L_\star \gtrsim 1$, which is energetically impossible unless the excess component is self-luminous (which is not normally the case for circumstellar material). Such a spectrum is however consistent with an ultraluminous infrared galaxy \citep[ULIRG, e.g.][]{2020NatAs...4..467A}. Thus, based on the spectrum alone, we conclude that it is most likely that the GDC is a background galaxy. We cannot however be definitive here because if the Keck source is extended, the flux could be higher, and thus the energetic argument weaker.

What is the likelihood that a background galaxy could be located directly behind the Fomalhaut debris disk, thus masquerading as a dust cloud or planet candidate? This question has been asked previously with ALMA observations of debris disks, albeit for much lower sensitivity data \citep[e.g.][]{2021MNRAS.506.1978L,2022MNRAS.517.2546L}. Following a similar methodology, using the 1.2\,mm sub-millimeter galaxy (SMG) counts of \citet{2020ApJ...897...91G}, we calculate the number of SMGs on the sky with a flux ${>}50\,\mu$Jy (i.e. $\approx$4$\sigma$) as approximately 37,000. Given the angular size of the Fomalhaut debris disk, i.e. an annulus with a radius $r=136$\,au and width $\Delta r=27$\,au inclined at 66$^\circ$, at its distance of 7.7\,pc we estimate that $0.5 \pm 0.05$ SMGs should appear behind the disk (accounting for the ALMA noise of $13\,\mu$Jy). While this same count rate would predict approximately ten sources appearing in entire ALMA image, it likely detects fewer because the ALMA sensitivity is not uniform across this region. This estimate suggests there is a fairly high probability that the ALMA source \almagdc~located just inside the dust ring is a background galaxy, and the same applies to the exterior sources.

Mid-IR galaxy counts at 21\,$\mu$m \citep[with JWST MIRI;][]{2022arXiv220901829W}, 24\,$\mu$m \citep[with Spitzer;][]{2004ApJS..154...70P}, and from models at 25.5\,$\mu$m \citep[with JWST MIRI;][]{2018MNRAS.474.2352C} predict similar number counts to ALMA; respectively 0.1--2, 6--8, and 4--8 for sources above 40$\mu$Jy in a 56$\times$56\,\arcsec~region on the sky (i.e. the MIRI FOV), and 0.01--0.2, 0.6--0.8, and 0.4--0.8 for the area of the disk annulus (where our threshold choice of 40$\mu$Jy corresponds to about 3$\times$ the RMS of the MIRI image). Empirically, the MIRI image suggests that the count is close to ten per FOV, in good agreement with \citet{2018MNRAS.474.2352C}, \citet{2004ApJS..154...70P} and the ALMA counts.
Although the counts are somewhat higher than \citet{2022arXiv220901829W}, the predictions of this latter study have large reported uncertainties, and their difference could be due to cosmic variance. 


Based on reasonable consistency with the expected number of detections from galaxy counts, and the total number of sources present in the images, we consider it probable that most, if not all, of the compact sources seen in the ALMA and JWST images are background galaxies. In the case of GDC, the flux levels at optical, mid-IR, and mm wavelengths, also argue in favor of a background galaxy.

\section{Conclusions}

By considering ALMA and Keck data, and the newly presented JWST MIRI data of Fomalhaut, we conclude that the source dubbed the ``Great Dust Cloud'' is a chance alignment with a background source. Each of the ALMA and Keck datasets independently detect multiple sources in common with JWST. For both ALMA and Keck a source is found at the sky location of the GDC, when an object that is co-moving with Fomalhaut would have moved 2-4\,\arcsec~between observations. Galaxy number counts suggest that at the high sensitivity provided by JWST, a coincidence such as a source like the GDC appearing within the disk was probable. The spectrum of the GDC also suggests that it is a background galaxy.

We also detect a number of other background sources, some of which are co-located with different Fomalhaut disk components as observed by JWST, and could influence geometric interpretations of Fomalhaut's disk. Given Fomalhaut's proper motion direction, many of the background sources that are not currently behind the disk will pass behind it in the future. Most of these are not apparent in the JWST images in Figure \ref{fig:jwst2325} so are unlikely to be a problem.

We find that there is a small astrometric shift between the 23 and 25.5\,$\mu$m images, with the former being 0.3\,\arcsec~to the east. This offset is visible for the GDC and other disk features.

More generally, we conclude that new direct imaging campaigns with JWST will be sufficiently sensitive to detect galaxies that are also visible in deep ($\sigma \lesssim 10$\,$\mu$Jy) ALMA observations, and that faint objects detected with JWST may also be detected with optical/near-IR high-contrast imaging. At least some JWST direct imaging programmes aim to detect planets in nearby systems that also host well-known disks, so it is possible that previously obtained ALMA data will help distinguish planets from background sources in other systems.

\section*{Acknowledgements}
We thank the referee for a succinct report.
GMK is supported by the Royal Society as a Royal Society University Research Fellow. JBL acknowledges support from the Smithsonian Institution as a Submillimeter Array (SMA) Fellow.
This paper makes use of the following ALMA data: ADS/JAO.ALMA\#2015.1.00966.S ALMA is a partnership of ESO (representing its member states), NSF (USA) and NINS (Japan), together with NRC (Canada), MOST and ASIAA (Taiwan), and KASI (Republic of Korea), in cooperation with the Republic of Chile. The Joint ALMA Observatory is operated by ESO, AUI/NRAO and NAOJ. This work is based on observations made with the NASA/ESA/CSA James Webb Space Telescope. These observations are associated with program 1193. Some of the data presented herein were obtained at the W. M. Keck Observatory, which is operated as a scientific partnership among the California Institute of Technology, the University of California and the National Aeronautics and Space Administration. The Observatory was made possible by the generous financial support of the W. M. Keck Foundation. The authors wish to recognize and acknowledge the very significant cultural role and reverence that the summit of Maunakea has always had within the indigenous Hawaiian community.  We are most fortunate to have the opportunity to conduct observations from this mountain.

\section*{Data Availability}

The ALMA data used here are available in the ALMA archive at \href{https://almascience.eso.org/aq}{https://almascience.eso.org/aq}. A script to split and image the ALMA data associated with \citet{2020RSOS....700063K} is at \href{https://github.com/drgmk/eccentric-width}{https://github.com/drgmk/eccentric-width}. The final JWST images associated with \citet{2023NatAs.tmp...93G} are available at \href{https://github.com/merope82/Fomalhaut}{https://github.com/merope82/Fomalhaut}. Keck data are available in the Keck Observatory Archive at \href{https://www2.keck.hawaii.edu/koa/public/koa.php}{https://www2.keck.hawaii.edu/koa/public/koa.php}.






\bsp	
\label{lastpage}
\end{document}